\documentstyle[12pt]{article}
\input BoxedEPS
\SetRokickiEPSFSpecial  
\HideDisplacementBoxes

\newcommand{\half}{\frac{1}{2}}
\begin{document}

\begin{center}

{\LARGE\bf Can Transversity Be Measured? }
\bigskip

{\large R.~L.~Jaffe}
\bigskip

{\it 
Center for Theoretical Physics,\\ 
Laboratory for Nuclear Science and
Department of Physics \\
Massachusetts Institute of Technology,
Cambridge, Massachusetts 02139
}\\
and\\
{\it RIKEN BNL Research Center\\
Brookhaven National Laboratory, Upton, NY, 11973\\[1ex]
\rm MIT-CTP\#2685\qquad hep-ph/9710465}
\bigskip

Invited paper, presented at the 2nd Topical Workshop,\\  DESY Zeuthen,
September 1--5, 1997:\\   ``Deep Inelastic Scattering off Polarized Targets:
Theory Meets Experiment"
\smallskip

\end{center}

\begin{abstract}\noindent
I review the ways that have been proposed to measure the quark
transversity distribution in the nucleon.  I then explain a proposal,
developed by Xuemin Jin, Jian Tang and myself, to measure transversity
through the final state interaction between two mesons ($\pi\pi$, $K
\overline K$, or $\pi K$) produced in the current fragmentation region
in deep inelastic scattering on a transversely polarized nucleon.
\end{abstract}

\section{Introduction}

\vspace{1mm}
\noindent
Three parton distributions characterize the quarks in the nucleon.  Two are 
now quite well known:  the momentum distribution, $q(x,Q^2)$,
 has been studied 
for decades \cite{rev1}; the helicity distribution, $\Delta q(x,Q^2)$, has 
been measured accurately only recently.\cite{rev2}  The discovery that the 
integrated quark helicity accounts for little of the nucleon's total spin 
initially shocked students of QCD and ignited a renaissance in QCD 
spin physics.

The third distribution has quite a different history.  It escaped
notice until 1979 when Ralston and Soper found it in their study of
Drell-Yan spin asymmetries.\cite{RS1979} Its place along with $q$ and
$\Delta q$ in the complete description of the nucleon spin was not
appreciated until after measurements of $\Delta q$ had spurred
interest in QCD spin physics.\cite{artru90,jaffe91,cortes92} It is known generally as the
``transversity distribution'', denoted $\delta
q(x,Q^2)$.\footnote{Although the name {\it transversity \/} is fairly
universal, the notation is not.  In addition to $\delta q$, the
notation $\Delta_T q$ due to Artru and Mekhfi\cite{artru90} is common,
as is the designation $\delta q \leftrightarrow h_1$ in analogy to
$\Delta q \leftrightarrow g_1$.}  $\delta q$ measures the distribution
of quark transverse spin in a nucleon polarized transverse to its (infinite) momentum.

Together, $q$, $\Delta q$, and $\delta q$ provide a complete description of 
 quark momentum and spin at leading twist.  This can be seen from a spin-density matrix 
representation of the leading twist quark distribution 
\begin{equation}
{\cal F}(x,Q^2) = {1\over 2} q(x,Q^2)~I\otimes I + {1\over 2} \Delta
q(x,Q^2)~\sigma_3 \otimes 
\sigma_3+{1\over 2}  \delta q(x,Q^2)~
\left(\sigma_+\otimes\sigma_-+\sigma_-\otimes\sigma_+\right) 
\label{symmetry}
\end{equation}
where the Pauli matrices lie in the quark and nucleon helicity spaces
respectively.  The completeness and complementarity of the three
distributions is obvious from the symmetry of eq.\thinspace (\ref{symmetry})
even without a detailed derivation and discussion.\cite{jaffe95}

The principal difference between transversity and the other two
distributions is that $\delta q$ is {\it systematically suppressed\/}
in deep inelastic lepton scattering (DIS) and other easily accessible
hard processes.  Theorists have proposed several experiments designed
to access $\delta q$, but up to now the prospects have not been
encouraging.  Help may be on the way, however.  I will describe a new
proposal to detect $\delta q$ by analysis of {\it two meson
fragmentation\/} in transversely polarized DIS.  This idea was
developed in collaboration with Xuemin Jin and Jian Tang and is
summarized in a recent preprint.\cite{JJT} We are optimistic that
careful study of $\pi^+\pi^-$, $K\pi$,  or $K \overline K$ final states will
yield a first look at this important quantity.

\section{Transversity}

\vspace{1mm}
\noindent

\begin{figure}[htbp]
$$
\BoxedEPSF{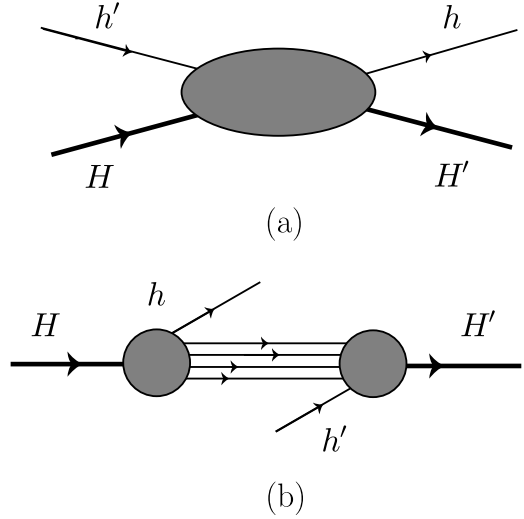 scaled 700}
$$
        \caption{Forward quark-hadron scattering.  (a) The forward
        scattering amplitude with helicities of quarks and hadrons
        labelled. (b) The particular ($u$-channel) discontinuity that
        corresponds to the quark distribution function.  It describes
        emission of a quark with helicity $h$ from a hadron of
        helicity $H$, and reabsorption of the quark with helicity $h'$
        to give a hadron of helicity $H'$.}
        \label{JaffeCTFig1:helicity}
\end{figure}

Quark distribution functions can be regarded as particular
discontinuities in forward quark-nucleon scattering.\cite{jaffe95}  Quarks are
represented by two-component spinors --- the ``good'' light-cone components
of the four-component Dirac fields.  The ``bad'' components enter at
non-leading twist.   A nucleon of
momentum $\vec P$ emits a quark of momentum fraction $x$ in the forward
direction.  Transverse momenta also enter only at higher twist.  The
emitted quark interacts in some hard process and then is reabsorbed by
the nucleon.  The whole process is collinear, so the angular momentum
along the $\hat P$-axis $\equiv$ helicity, is conserved.  The
scattering amplitude and the discontinuity that defines a distribution
function are shown in Fig.~[\ref{JaffeCTFig1:helicity}].
%
%
%
%

The quark ($h$, $h'$) and
hadron ($H$, $H'$) helicities take on the values $\pm 1/2$.
Helicity conservation requires $H+h'=H'+h$.  Parity sends
$h\rightarrow-h$, {\it etc.\/}, and time-reversal interchanges initial
($H,h'$) and final ($H',h$) helicities.  This leaves exactly three
independent quark distribution functions corresponding to the helicity
labels,
\begin{equation}\def\half{\displaystyle{\frac{1}{2}}}
\begin{tabular}{rrcrr}
$\half$ & $\half$ & $\phantom{-}\longrightarrow$ & $\half$ & $\half$\\[2ex]
$\half$ & $-\half$ & $\phantom{-}\longrightarrow$ & $\half$ & $-\half$\\[2ex]
$\half$ & $-\half$ & $\phantom{-}\longrightarrow$ & $-\half$ & $\half$ 
\end{tabular}\label{helamp}
\end{equation}
in the notation $H\:h'\rightarrow H'\:h$. These three helicity
amplitudes can be regrouped into combinations that measure spin
average ($q$), helicity difference ($\Delta q$), and helicity flip
$\equiv$ transversity ($\delta q$) distribution,
\begin{equation} \def\half{\displaystyle{\frac{1}{2}}}
\begin{tabular}{ccl}
    $q$ & $\longleftrightarrow$ & 
   $\Bigl(\begin{tabular}{rrcrr}
          $\half$ & $\half$ & $\rightarrow$ & $\half$ & $\half$
             \end{tabular}\Bigr) + 
    \Bigl(\begin{tabular}{rrcrr}
       $\half$ & $-\half$ & $\rightarrow$ & $\half$ & $-\half$
        \end{tabular} \Bigr)$ \\[2ex]
   $ \Delta q$ & $\longleftrightarrow$ & 
   $\Bigl(\begin{tabular}{rrcrr}
   $\half$ & $\half$ & $\rightarrow$ & $\half$ & $\half$
   \end{tabular}\Bigr) - \Bigl(
   \begin{tabular}{rrcrr}
   $\half$ & $-\half$ & $\rightarrow$ & $\half$ & $-\half$
   \end{tabular} \Bigr)$ \\[2ex]
   $ \delta q$ & $\longleftrightarrow$ & 
   $\Bigl(\begin{tabular}{rrcrr}
   $\half$ & $-\half$ & $\rightarrow$ & $-\half$ & $\half$
   \end{tabular}\Bigr)$
\end{tabular}\label{qs}
\end{equation}
When this result is 
written in terms of $2\times 2$ matrices in the helicity spaces
of the quark and the nucleon, we obtain the elegant and simple matrix
structure of eq.\thinspace (\ref{symmetry}).

The distributions $q$ and $\Delta q$ are the sum and difference,
respectively, of the probability
to find a quark polarized along and against the
polarization of a nucleon in a helicity eigenstate.  The transversity
is off-diagonal in helicity and does not appear to have a
probabilistic interpretation.  This is an accident of working in a
helicity basis.  If, instead, we work in a basis of transverse spin
eigenstates, then the transversity is a difference of probabilities
and $\Delta q$ is off-diagonal.  In short,
\begin{eqnarray}
   \Delta q_a(x,Q^2) &=& q_a^\uparrow(x,Q^2) -
   q_a^\downarrow(x,Q^2)\nonumber\\[1ex]
   \delta q_a(x,Q^2) &=& q_a^\bot(x,Q^2) -
   q_a^\top(x,Q^2)\ .\label{deltas}
   \end{eqnarray}

\begin{figure}[htbp]
$$
\BoxedEPSF{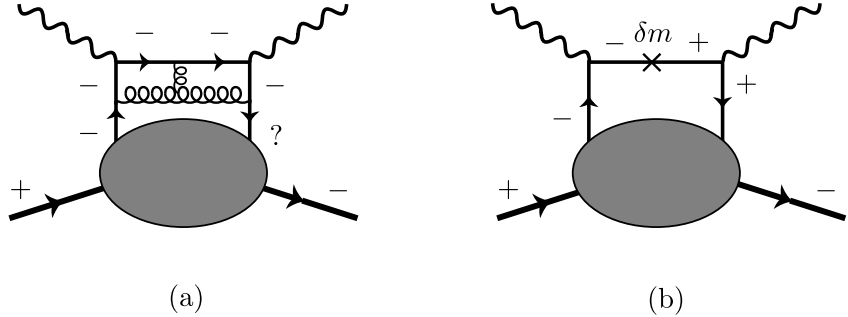 scaled 900}
$$
        \caption{Why $\delta q$ decouples from DIS.  (a) A typical
        perturbative contribution to DIS.  The chirality of the
        propagating quark is not changed by the coupling to gluons or
        photon, so it cannot be reabsorbed to form the outgoing
        hadron.  (b) A mass insertion (marked with an $\times$) can
        flip chirality but give a contribution suppressed by ${\cal
        O}(1/Q^2)$.}
        \label{JaffeCTFig2:chiralflip}
\end{figure}

 The differences between $\Delta q$ and $\delta q$ begin to
emerge when we look at their
chirality properties.  Helicity and chirality are identical for
``good'' light-cone components of the Dirac field (they are opposite
for the ``bad'' components).  So the quark-hadron forward scattering
amplitude that defines $\delta q$ flips the chirality as well as the
helicity of the quark.  Since all perturbative QCD process and all
interactions with external electroweak currents conserve chirality (up
to corrections of order $m/Q$), $\delta q$ decouples from all hard
processes that involve only one quark distribution (or fragmentation)
function.  The decoupling of such ``chiral-odd'' distributions is
illustrated in Fig.~[\ref{JaffeCTFig2:chiralflip}].  So it is impossible to 
%
%
measure transversity in ordinary deep inelastic scattering,
$ep\rightarrow e'X$, or single particle inclusive annihilation,
$e^+e^-\rightarrow h X$.  In the next section I will explore processes
where transversity does couple.

In the preceeding talk at this workshop, Jacques Soffer has reviewed
many of the properties of the transversity, and its lowest moment, the
tensor charge.\cite{Soffer}  Here is a brief summary for the sake of
completeness.
\begin{itemize}
    \item For non-relativistic quarks, where boosts and rotations
    commute, $\Delta q_a(x,Q^2)=\delta q_a(x,Q^2)$.  However, $g_A/g_V=5/3$ and
    $\Delta q = 1$ for non-relativistic quarks, and both relations
    fail in reality ($g_A/g_V\approx 1.26$ and $\Delta q \approx 0.2$).
    The quarks inside the nucleon cannot be
    non-relativistic, and the difference between the nucleon's
    helicity and transversity distributions provides a further 
    and more detailed measure
    of the relativistic nature of the quarks inside the nucleon.

    \item The lowest moment of $\delta q$ measures a simple local
    operator analogous to the axial charge,
    \begin{eqnarray}
    \int_{-1}^1 dx\, \delta q_a(x,Q^2)& =& \int_0^1 dx\,  (\delta q_a(x,Q^2) -
    \delta \bar q_a(x,Q^2)) \equiv \delta q_a(Q^2)\nonumber\\
    &\sim& \langle PS_\perp | \bar q_a \sigma^{0\perp}\gamma_5 q_a\vert_{Q^2} |
    PS_\perp\rangle
    \label{tensor}
    \end{eqnarray}
    known as the ``tensor charge''.  Unlike its vector and axial
    equivalents, the tensor charge is not conserved in QCD, so it has
    an anomalous dimension at one-loop.  One of the outstanding
    puzzles in this field is how to obtain an independent measure of
    $\delta q_a(Q^2)$ and thereby formulate a ``transversity sum
    rule'' analogous to those that have been so helpful in the study
    of $\Delta q_a$.
    
    \item $\delta q(x,Q^2)$ does not mix with gluons under evolution.
    All its anomalous dimensions are positive.  Thus it evolves very
    simply with $Q^2$ --- it goes slowly and inexorably to zero.  Even
    its area, which corresponds to the tensor charge, is not
    conserved.  The anomalous dimensions for $\delta q$ were first
    calculated by Artru and Mekhfi\cite{artru90} in leading order.
    NLO anomalous dimensions have now been calculated by three
    groups\cite{andim} and are reported here by Koike.\cite{Koike}
    $\delta q(x,Q^2)$ has been studied in hadron
    models.\cite{jaffe91,models}

    \item Initial efforts have been made to calculate the tensor charge on 
    the lattice.\cite{lattice}  Although the lattice is small
    ($16^3\times 20$) and the coupling is relatively strong ($\beta =
    5.7$), the results are intriguing:  $\delta u + \delta d + \delta
    s = 0.562 \pm 0.088$, to be compared with the experimental
    result $\Delta u + \Delta d + \Delta s \approx 0.2$.  The tensor
    charge seems to behave much more like the ``naive'' quark model
    expectation that the quarks carry much of the nucleon's spin.

    \item The transversity obeys some important inequalities.  The
    first, $|\delta q_a(x,Q^2)|\le q_a(x,Q^2)$, follows from its
    interpretation as a difference of probabilities.  The second has
    its origins in the positivity properties of helicity amplitudes,
    $|2\delta q_a(x,Q^2)|\linebreak
 \le q_a(x,Q^2) + \Delta
    q_a(x,Q^2)$.\cite{Soffer,Soffer2} Saturation of Soffer's
    inequality along with simple model calculations would
    suggest $|\delta q_a|> |\Delta q_a|$.  Although transversity is
    hard to measure, there is no indication that it is small.

\end{itemize}

\section{Proposals to Measure Transversity}

\begin{figure}[htbp]
$$
\BoxedEPSF{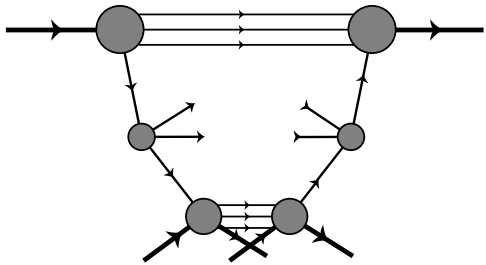 scaled 900}
$$
        \caption{A generic process that could be sensitive to the
        nucleon's transversity.  The presence of {\it two\/} soft
        processes is crucial since the chirality flip associated with
        $\delta q$ must be reversed by a second soft interaction.}
        \label{JaffeCTFig3:twoflip}
\end{figure}

\vspace{1mm}
\noindent
Coupling to the transversity requires a chirality flip
that perturbative QCD cannot provide.  A
second soft hadronic vertex must be present if transversity
is to figure in a hard QCD process.  A generic (and complicated)
example is shown in Fig.~[\ref{JaffeCTFig3:twoflip}].
%
%
%

The uppermost soft vertex might be the transversity distribution of
the target.  The compensating chirality flip can be provided by a
distribution, fragmentation, or fracture function at the lower
vertex.  In the figure a fracture function describing beam
fragmentation does the job.  Limiting ourselves to distribution and
fragmentation functions\footnote{To my knowledge the helicity and
chirality analysis of fracture functions has yet to be attempted.} we
can catalog the available chiral-odd ({\it i.e.\/} chirality flip)
quantities: 
\begin{itemize}

    \item At leading twist (${\cal O}(Q^0)$), the transversity,
    $\delta q(x,Q^2)$ and the fragmentation function with the same helicity
    structure, $\delta\hat q(z,Q^2)$, are the only chirality flip quantities.

    \item At twist three (${\cal O}(\Lambda/Q)$), we have
    \begin{itemize}
      \item  a spin-average
    distribution function, $e(x,Q^2)$, associated with the operator
    $\bar qq$, and the fragmentation function with the same helicity
    structure, $\hat e(z,Q^2)$; and 
      \item  a helicity-difference distribution
    function, $h_L(x,Q^2)$ and its fragmentation analog, $\hat
    h_L(z,Q^2)$. 
    \end{itemize}
     For more information on these unfamiliar distributions, see
    Ref.~\cite{JafJi}.
\end{itemize}

With these building blocks, it is possible to critique the existing
proposals to measure $\delta q$.
\begin{itemize}
    \item  Transversely Polarized Drell-Yan in $pp$ Collisions.\cite{RS1979}
    \begin{itemize}
        \item  Process: $\vec p_\perp \vec p_\perp 
        \rightarrow \ell\bar\ell X$
        \item  Measures:  $\delta q(x_1)\otimes\delta\bar q(x_2)$
        \item  Advantages:
	    \begin{itemize}
            \item Drell-Yan is well understood through next to leading
   		 order in perturbative QCD.\cite{Kamal}
            \item The RHIC Collider in its polarized mode will
            provide the beams.
            \end{itemize}
        \item  Disadvantages:
            \begin{itemize}
	    \item $\delta\bar q$ is presumably small in the nucleon.
            \item Drell-Yan rates are generically low  compared to
    		purely hadronic processes at colliders.
            \end{itemize}
    \end{itemize}
    \item Twist-three Pion Production in DIS.\cite{JafJi}
    \begin{itemize}
        \item  Process: $e \vec p_\perp 
        \rightarrow e' \pi X$
        \item  Measures:  $\delta q(x)\otimes \hat e(z)\:\:\oplus\:\:
        g_T(x)\otimes \hat q(z)$
        \item  Advantages:
	    \begin{itemize}
            \item Pions are abundant in the fragmentation region of DIS.
            \item Accessible to HERMES or COMPASS with a transversely
            polarized target.             
            \end{itemize}
        \item  Disadvantages:
            \begin{itemize}
	    \item The process is twist-three,
            suppressed by ${\cal O}(1/Q)$.
            \item The competing (twist-three) chiral-even mechanism,
            $g_T(x)\otimes \hat q(z)$, must be subtracted away to
            expose the process of interest.
            \end{itemize}
    \end{itemize}
 \item $A_{TT}/A_{LL}$ in Polarized Jet Production.\cite{Ji,JafSaito,Barone}
    \begin{itemize}
        \item Process:  $p_\perp p_\perp \rightarrow {\rm jet(s)} X$.
        \item  Measures:
		\begin{equation}
		\frac{\delta q(x_1)\otimes \delta
   		 q(x_2)}{\Delta G(x_1)\otimes \Delta
		G(x_2) \oplus \Delta G(x_1) \otimes \Delta q(x_2) + \cdots}
		\nonumber
		\end{equation}
        \item  Advantages:
	    \begin{itemize}
            \item Jets are abundant in the final state at colliders.
            \item Accessible to RHIC in the polarized collider mode.          
            \end{itemize}
        \item  Disadvantages:
            \begin{itemize}
	    \item The production of jets in collisions of transversely
    		    polarized protons is suppressed for a variety of
    		    reasons.  Most important is the fact that there is
    		    no analog of transversity for gluons.  Furthermore
		    the quark-quark contribution is suppressed by a
		    color exchange factor.\cite{Ji}
		In fact, the suppression of ${\cal A}_{TT}$
		relative to ${\cal A}_{LL}$ is so enormous that it
		can be viewed as a qualitative test of the gauge 
		nature of gluons in QCD.\cite{JafSaito}
	
            \end{itemize}
    \end{itemize}
\end{itemize}    

A suggestion by Collins and collaborators\cite{collins} requires
special consideration because of its close relation to the new proposal
described in the following section.  They proposed to look for an
asymmetry associated with a plane defined in the current
fragmentation region in DIS.  There are many ways in which the current
fragmentation region might define a plane.  Since the virtual photon
momentum, $\vec q$, defines the fragmentation axis, any other vector,
such as the momentum of a meson, perhaps the leading meson, or some
event shape measure like oblateness, serves to define a second vector,
$\vec p$.  Then the correlation of the normal to the plane formed by
$\vec q$ and $\vec p$ with the transverse spin of the target,
  \begin{equation}
    \cos\phi = \frac{\vec p\times\vec q\cdot\vec S_\perp}{|\vec p\times
    \vec q||\vec S_\perp|}
  \end{equation}
is sensitive to the transversity of the struck quark.  $\phi$ has
become known as the ``Collins angle''.

Application of ``naive time-reversal'' symmetry,\cite{Conzett} ($\vec
p$, $\vec q$, $\vec s$)$ \rightarrow $ $-$ ($\vec p$, $\vec q$, $\vec
s$), requires $\langle \cos\phi \rangle = 0$ and seems to forbid a
Collins angle asymmetry.  This can be
avoided if there are significant phases generated by final state
interactions (FSI) in the fragmentation process.  Although phases are
expected in general, the exact phase presumably depends sensitively on
the kinematic relation between the observed particle (or particles),
$\vec p$, and the remainder of the current jet, $X$.  It seems quite
likely that the sum over unobserved particles, $X$, would tend to
average the phase to zero and suppress the effect.  Up to now,
searches for a dependence of fragmentation on Collins angle and
similar effects like ``handedness'' have come up negative.\cite{zeuthen}
With these comments in mind we can summarize,
\begin{itemize}
 \item Observation of a Collins angle dependence in final state
    fragmentation.\cite{collins}
    \begin{itemize}
        \item  Process: $e \vec p_\perp 
        \rightarrow e' X$
        \item  Measures:  $\delta q(x)\otimes\delta \hat q(z)$
        \item  Advantages:
	    \begin{itemize}
            \item Every DIS event has a current fragmentation region.
            \item Accessible to HERMES or COMPASS running with a transversely
            polarized target.             
            \end{itemize}
        \item  Disadvantages:
            \begin{itemize}
	    \item Requires a FSI phase that does not average to zero
    		summed over unobserved hadrons, $X$.
             \end{itemize}
    \end{itemize}
\end{itemize}

\section {Transversity and Two-Particle Fragmentation}

\vspace{1mm}
\noindent

Recently Xuemin Jin, Jian Tang and I\cite{JJT} have studied a way to
measure transversity that may be free from many of these
shortcomings. The process of interest is two-meson ({\it e.g.\/}
$\pi^+\pi^-$, $\pi K$, or $K\overline K$) production in the current
fragmentation region in deep inelastic scattering on a transversely
polarized nucleon. Our analysis focuses on the {\it interference}
between the $s$- and $p$-waves of the two-meson system around the
$\rho$ (for pions), the $K^*$ (for $K\pi$), or the $\phi$ (for
kaons). Such an interference effect allows the quark's polarization
information to be carried through a dependence of the cross section on
    \begin{equation}
    \vec k_+ \times \vec k_- \cdot  \vec S_\perp 
    \end{equation}
  where
$\vec k_+$, $\vec k_-$, and $\vec S_\perp$ are the three-momenta of 
$\pi^+$ ($K$), $\pi^-$ ($\overline K$), and the nucleon's
transverse spin, respectively. 

 This effect is at the leading twist level, and the production rates
for pions and kaons are large in the current fragmentation region.
However, it would vanish by T-invariance in the  absence of final
state interactions, or by C-invariance if the two-meson state were an
eigenstate of C-parity.   Both suppressions are evaded in the $\rho$
($\pi^+\pi^-$), $\phi$ ($K\overline K$), or $K^*$ ($K\pi$) mass regions.

The final state interactions of $\pi\pi$, $\pi K$, and $ K\overline{K}$ are
known in terms of  meson-meson phase shifts. From these phase shifts
we know that $s$- and $p$-wave production channels interfere strongly
in the mass region around the $\rho$, $K^*$, and $\phi$ meson resonances.
Since the $s$- and $p$-waves have opposite C-parity, the interference
provides exactly the charge conjugation mixing necessary.  Combining
perturbative QCD, final state interaction theory, and data on the
meson-meson phase shifts, we can relate this  asymmetry to known
quantities, the transversity distribution we seek, and to a new  type
of fragmentation function that describes the $s$- and $p$-wave
interference  in the process $q\rightarrow \pi^+\pi^-$
($\pi K$, $K\overline{K}$). Unless this fragmentation is  anomalously
small, the measurement of this asymmetry would allow  extraction of
a (charge-squared weighted) quark transversity distribution.

\begin{figure}[htbp]
$$
\BoxedEPSF{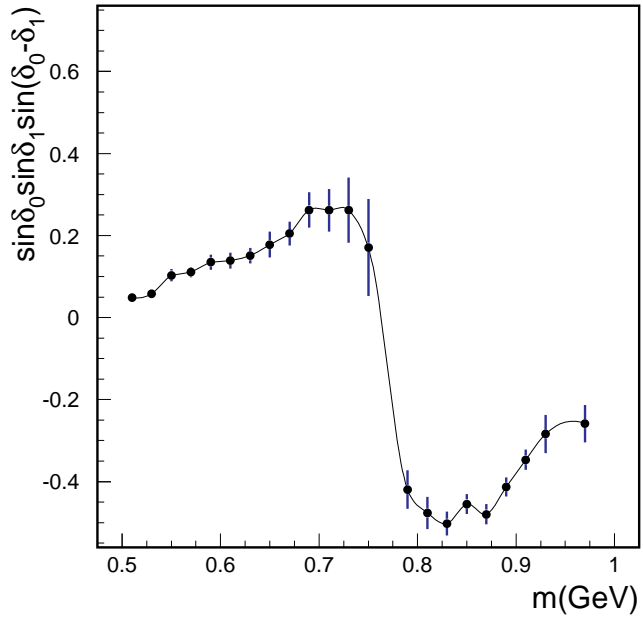 scaled 600}
$$
        \caption{The factor, $\sin\delta_0 \sin\delta_1
\sin(\delta_0-\delta_1)$, 
           as a function of the invariant mass $m$ of two-pion
system. 
           The data on $\pi\pi$ phase shifts are taken from
Ref.~\protect\cite{ref1}.}
        \label{JaffeCTFig4:fig2}
\end{figure}

 In the style of the previous section we summarize the characteristics
of this process:
\begin{itemize}
 \item Two Meson Correlation with Transverse Spin \cite{JJT}
    \begin{itemize}  
	\item  Process: $e  \vec p_\perp  \rightarrow e'
             \pi^+\pi^- X $.  Generalizations to other meson pairs and
             other processes like
             $p\vec p_\perp\rightarrow \pi^+\pi^- X$, {\it etc.\/}, are
             possible.
        \item  Measures:  $\delta q(x_1)\otimes\delta\hat q_I(z)$,
            where the latter is an {\it interference fragmentation 
            function\/}.
        \item  Advantages:
	    \begin{itemize}
	    \item The effect is leading twist.
            \item Meson pairs are copiously produced in quark fragmentation. 
            \item The meson-meson final state interaction phase
   		 remains fixed as the sum on $X$ is performed.
	    \item The final state interaction phase can be measured 
		 in meson-meson scattering.
	    \item The HERMES and (in the future) COMPASS
		 can perform the measurement.
            \end{itemize}
        \item  Disadvantages:
            \begin{itemize}
	    \item The cross-section must be held differential to avoid
		 averaging the phase to zero (see below).
            \end{itemize}
    \end{itemize}
\end{itemize}
Some of the subtleties of this analysis can be addressed by answering
some obvious questions:
\begin{itemize}

	\item Doesn't $\langle \vec \pi^+ \times \vec \pi^-\cdot \vec
		S_\perp\rangle$ vanish by charge conjugation ($\pi^+
		\leftrightarrow \pi^-$)?

	Answer:  Only if $ \pi^+\pi^- $ is in a C-conjugation
	eigenstate.  So if the two pions form a $\rho$ meson ($C=-1$)
	there is no effect.  Since $C=(-1)^\ell$ for $\pi\pi$, more
	than one partial wave must be active to generate an effect.

	\item Why don't phases drop out of the sum over final states
		of the  quantity
		\begin{equation}
		     \sum_X |\pi^+\pi^-X\rangle_{\rm out}\:_{\rm out}
		     \langle\pi^+\pi^-X|\ ?
		\end{equation}
        since whatever phase occurs in the {\it bra\/} also occurs in
		the {\it ket\/}?

	Answer: This would indeed be the case if the final state had a
		 unique phase.  However, if $(\pi^+\pi^-)$ is a {\it
		 superposition\/} of two or more different partial
		 waves with different relative phases,  {\it e.g.\/}
		\begin{equation}
		     |\pi^+\pi^-X\rangle_{\rm out} = e^{i\delta_0}
		 |(\pi\pi)^{\ell=0}X\rangle + e^{i\delta_1}
		 |(\pi\pi)^{\ell=1}X\rangle + \cdots 
		\end{equation}
		then the interference between $s$ and $p$
		 waves, which has a unique signature in the angular
		 distribution, is proportional to the (imaginary part
		 of) the phase $e^{i(\delta_0-\delta_1)}$.
\end{itemize}
So we conclude that interfering $s$ and $p$-waves in $\pi^+\pi^-$
 ($K\overline K$, or $K\pi$) production is an excellent way to satisfy
 all these symmetry constraints and still allow a $\vec k_+ \times
 \vec k_- \cdot \vec S_\perp$ signature.  The ``figure of merit'' for
 the necessary final state interactions is the product of the $s$-wave
 amplitude, proportional to $\sin\delta_0$, the $p$-wave amplitude,
 proportional to $\sin\delta_1$, and the interference, proportional to
 $\sin(\delta_0-\delta_1)$.  Good data on $\pi\pi$ phase shifts are
 available from the 1960's and 1970's.\cite{ref1} The figure of merit
 is plotted in Fig.~\ref{JaffeCTFig4:fig2}.  
%
%
It is interesting to note that the maximum possible value of the 
product of these three sines is $3\sqrt{3}/8$.  The $\pi\pi$ FSI's
fortuitously come within $75\%$ of this value.

  Our work can be
viewed as a
 specialization of the Collins effect to the angle defined by the
 $\pi^+\pi^-$, $K\pi$, or $K\overline K$ final state.  Several authors
 \cite{collins,ji94,collins2} have mentioned or considered a
 $\vec\pi^+\times\vec\pi^-\cdot\vec S_\perp$ correlation in DIS but
 not explored it in depth.\footnote{An exception is
 Ref.~\cite{collins2} where the authors study the Collins angle in
 the linear sigma model.  The angular correlation arises through the
 interference of the $\pi\pi$ decay of a scalar, Breit-Wigner $\sigma$
 resonance and the $\pi\pi$ continuum generated by successive pion
 emission from a fragmenting quark line. Although this demonstrates
 the existence of an asymmetry, it does not model the actual $\pi\pi$
 final state interaction accurately.}

The analysis of two particle fragmentation is rather new.  It was
first addressed in Ref.~\cite{collins2}.  We have developed a complete
formalism\cite{ref2}, but give only a brief introduction here,
restricted to the interference terms that contribute to this
asymmetry.  We consider the semi-inclusive deep inelastic scattering
process with two-pion final states being detected: $eN\rightarrow
e'\pi^+\pi^- X$.  The nucleon target is transversely polarized with
polarization vector $S_\mu$.  We adopt the standard variables in DIS,
$x = Q^2/2P\cdot q$ and $y = P\cdot q/P\cdot k$.  The $\sigma
[(\pi\pi)^{I=0}_{l=0}]$ and $\rho [(\pi\pi)^{I=1}_{l=1}]$ resonances
are produced in the current fragmentation region with momentum 
$P_h^\mu$
and momentum fraction $z = P_h\cdot q/q^2$. We recognize that the
$\pi\pi$ s-wave is not resonant in the vicinity of the $\rho$ and our
analysis does not depend on a resonance approximation.  For simplicity
we refer to the non-resonant s-wave as the ``$\sigma$''.  The
invariant squared mass of the two-pion system is $m^2 = (k_++k_-)^2$,
with $k_+$ and $k_-$ the momentum of $\pi^+$ and $\pi^-$,
respectively. The decay polar angle in the rest frame of the two-meson
system is denoted by $\Theta$, and the azimuthal angle $\phi$ is
defined as the angle of the normal of two-pion plane with respect to
the polarization vector $\vec S_\perp$ of the nucleon, $\cos\phi =
{{\vec k_+}\times{\vec k_-}\cdot\vec S_\perp / |\vec k_+\times \vec
k_-||\vec S_\perp|}$.

To simplify our analysis we make a collinear approximation in
referring the fragmentation coordinate system to the axis defined by
the incident electron.  At SLAC, HERMES, and COMPASS energies, a
typical value for the electron scattering angle is less than $0.1$.
Complexities in the analysis of fragmentation turn out to be
proportional to $\sin^2\theta$ and can be ignored at fixed target
facilities of interest. In this approximation the production of two
pions can be viewed as a collinear process with the electron beam
defining the common $\hat{e}_3$ axis. Also we take $\vec S_\perp$
along the $\hat{e}_1$ axis.

\begin{figure}[htbp]
$$
\BoxedEPSF{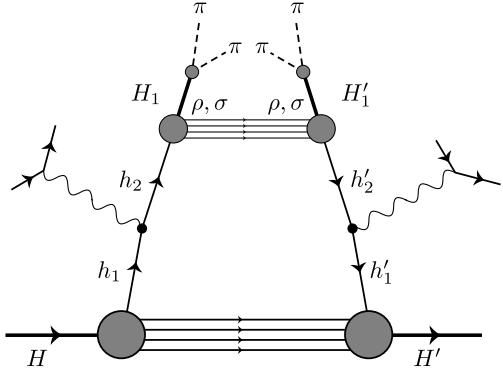 scaled 900}
$$
        \caption{Hard scattering diagram for $\pi^+\pi^-
(K\overline{K})$ production in
the current fragmentation region of electron scattering from a target
nucleon. In perturbative QCD the diagram (from bottom to top) factors
into the products of distribution function, hard scattering,
fragmentation function, and final state interaction. Helicity density matrix
labels are shown explicitly.}
        \label{JaffeCTFig5:fig1}
\end{figure}

Since we are only interested in a result at the leading twist, we
follow the helicity density matrix formalism developed in
Refs.~\cite{ref3,ref4}, in which all spin dependence is
summarized in a {\it double} helicity density matrix. We factor the
process at hand into basic ingredients: the $N\rightarrow q$
distribution function, the hard partonic $eq\rightarrow e'q'$ cross
section, the $q \rightarrow (\sigma, \rho)$ fragmentation, and the
decay $(\sigma, \rho)\rightarrow \pi\pi$, all as density matrices in
helicity basis:
\begin{eqnarray}
\left[{{d^6\sigma}\over{dx\, dy\, dz\, dm^2\, d\cos\Theta\,
d\phi}}\right]_{H'H}&& \nonumber \\ &&\hspace*{-2.8cm}=\left[{\cal
F}\right]\,_{H'H}^{h_1h'_1}\left[{{d^2\sigma(e q\rightarrow e' q')}
\over{dx\, dy}}\right]_{h'_1h_1}^{h_2h'_2} \left[{{d^2\hat{\cal
M}}\over{dz\, dm^2}}\right]_{h'_2h_2}^{H_1H'_1} \left[{{d^2{\cal
D}}\over{d\cos\Theta\, d\phi}}\right]_{H'_1H_1}
\label{hme} 
\end{eqnarray}
where $h_i(h_i')$ and $H(H')$ are indices labeling the helicity states
of quark and nucleon, and $H_1(H'_1)$ labeling the helicity state of
resonance the ($\sigma$, $\rho$). See Fig.~\ref{JaffeCTFig5:fig1}.
%
%

 In
order to incorporate the final state interaction, we have separated the
$q\rightarrow \pi^+\pi^-$ fragmentation process into two steps. First,
the quark fragments into the resonance ($\sigma$, $\rho$), then the
resonance decays into two pions, as shown at the top part of the
Fig.~\ref{JaffeCTFig5:fig1}.

The two meson fragmentation function depends on the momentum fraction
of the two pion system, $z$, the $\pi\pi$ invariant mass, $m$, and the
decay angles $\Theta$ and $\phi$.  Two types of final state
interactions can generate the necessary final state phase: i) those
between the two pions, and ii) those between the pions and the
hadronic state $X$.  We ignore the latter because we expect the phase
to average to zero when the sum on $X$ is performed ---
$|\pi^+\pi^-X\rangle_{\rm out}\rightarrow |(\pi^+\pi^-)_{\rm
out}X\rangle$.  The final
state phase of the two-pion system is determined by the $\pi\pi\:
{\cal T}$-matrix \cite{ref5}.  We separate out the phase for
later consideration and analyze the (real) $\rho$-$\sigma$
interference fragmentation function as if the two particles were
stable.

The $s$-$p$ interference fragmentation function describes the emission
of a $\rho (\sigma)$ with helicity $H_1$ from a quark of helicity
$h_2$, followed by absorption of $\sigma (\rho)$, with helicity $H'_1$
forming a quark of helicity $h'_2$. Conservation of angular momentum
along the $\hat{e}_3$ axis requires
\begin{equation}
H_1 + h_2' = H_1' + h_2\ .
\label{conservation}
\end{equation}
Parity and time reversal restrict the number of independent components  of
$\hat{\cal M}$:
\begin{eqnarray}
\hat{\cal M}^{sp(ps)}_{H_1'H_1,h_2h'_2} & = &
-\hat{\cal M}^{sp(ps)}_{-H_1'-H_1,-h_2-h'_2}\quad {\rm 
	(parity)} \label{parity}\\
\hat{\cal M}^{sp}_{H_1'H_1,h_2h'_2}& = &
\hat{\cal M}^{ps}_{H_1H_1',h'_2h_2}\quad {\rm 
	(T-reversal)}
\label{pt}
\end{eqnarray}
where the minus sign in eq.\thinspace (\ref{parity}) accounts for the parity
difference between $s$- and $p$-wave states. Note that eq.\thinspace (\ref{pt})
holds only after the T-reversal violating final state interaction
between two pions is separated out.  After these symmetry
restrictions, only two independent components remain, $\hat{\cal
M}^{sp}_{00,++}$ and $\hat{\cal M}^{sp}_{01,+-}$, and they can be
identified with two novel interference fragmentation functions,
$\Delta \hat q_{_I}$ and $\delta \hat q_{_I}$, respectively, where the
subscript $I$ stands for interference. Here, to preserve clarity, the
flavor, $Q^2$, and $z$ labels are  suppressed.  The helicity
$\pm\frac{1}{2}$ states of quarks are denoted $\pm$, respectively.
Hermiticity and time reversal invariance guarantee that $\Delta \hat
q_{_I}$ and $\delta \hat q_{_I}$ are real.  This information can be encoded into a double density matrix,
${d^2\hat{\cal M}}\over{dz\, dm^2}$ that carries the helicity indices
of the fragmenting quark and the meson resonance,
$\{\sigma,\vec\rho\}$.  For explicit constructions see
Ref.~\cite{JJT}. 

From its definition in terms of $\hat{\cal M}^{sp}_{01,+-}$ it is
clear that the interference fragmentation function, $\delta
\hat{q}_{_I}$, is associated with quark helicity flip and is therefore
chiral-odd. It is this feature that enables us to access the
chiral-odd quark transversity distribution in DIS.

The final state interactions between the two pions are  included
explicitly in
\begin{equation}
\Delta_0(m^2)=-i \sin\delta_0
e^{i\delta_0}\ ,\hspace*{1cm} 
\Delta_1(m^2)=-i \sin\delta_1
e^{i\delta_1} 
\label{propagators}
\end{equation}
where $\delta_0$ and $\delta_1$ are the strong interaction $\pi\pi$
phase shifts.  Here we have suppressed the $m^2$ dependence of the
phase shifts for simplicity.

The more conventional parts of the calculation are the quark
distribution function, the hard scattering process, and the
$\{\rho,\sigma\}$ decay density matrix.  All can 
be found in Ref.~\cite{JJT}.

Combining all the ingredients together, we obtain the polarization
dependent part of the cross section for the production of two pions
(kaons) in the current fragmentation region for unpolarized electrons
incident on a transversely polarized nucleon as follows
\begin{eqnarray}
{{d^6\sigma_\bot}\over{dx\, dy\, dz\, dm^2\, d\cos\Theta\,
d\phi}}
&=& \frac{e^4}{32\pi^3 Q^2 m}\, { 1-y\over y}
\sqrt{6}\, \sin{\Theta}\cos{\phi}\,\,
\nonumber
\\*[7.2pt]
&&\times
\sin{\delta_0} \sin{\delta_1} 
\sin\left({\delta_0-\delta_1}\right)\,
\sum_a e_a^2 \delta q_a(x)\, \delta \hat{q}_{_I}^a(z)\ .
\label{cross-section}
\end{eqnarray}
Here the sum over $a$ covers all quark and antiquark flavors.

An asymmetry is obtained by dividing out the polarization independent
cross section,
\begin{eqnarray}
{\cal A}_{\bot\top}\equiv {d\sigma_\bot -d\sigma_\top
\over d\sigma_\bot +d\sigma_\top}
&=& {\sqrt{6}(1-y)\over 1+(1-y)^2}\, \sin{\Theta}\cos{\phi}\,\,
\sin{\delta_0} \sin{\delta_1}
\sin\left(\delta_0-\delta_1\right) 
\nonumber
\\*[7.2pt]
&&\hspace*{1cm}\times 
{\sum_a e_a^2 \delta q_a(x)\, \delta \hat{q}_{_I}^a(z)\over
\sum_a e^2_a q_a(x)
\left[ \sin^2\delta_0
\hat{q}_0^a(z)
+\sin^2\delta_1
\hat{q}_1^a(z)\right]} 
\label{asymmetry}
\end{eqnarray}
where $\hat{q}_0$ and $\hat{q}_1$ are spin-average fragmentation functions
for the $\sigma$ and $\rho$ resonances, respectively. This asymmetry can be
measured either by flipping the target transverse spin or by binning events
according to the sign of the crucial azimuthal angle $\phi$. Note that this
asymmetry only requires a transversely  polarized nucleon target, but not a
polarized electron beam.

A few comments can be made about our results. The final state phase
generated by the $s$-$p$ interference is crucial to this analysis. If
the data are not kept differential in enough kinematic variables, the
effect will almost certainly average to zero.  We are particularly
concerned about the two-meson invariant mass, $m$, where we can see
explicitly that the interference averages to zero over the $\rho$ as
shown in Fig.~\ref{JaffeCTFig4:fig2}. It is unclear at this stage whether the
effect would survive averaging over the $z$ dependence of the
interference fragmentation function.

Finally, let us  examine the flavor content of the asymmetry. 
For $\pi^+\pi^-$ production, isospin symmetry gives
$\delta\hat{u}_{_I}
= - \delta\hat{d}_{_I}$ and $\delta\hat{s}_{_I} = 0$.
Charge conjugation implies $\delta \hat{q}^a_{_I}=-\delta \hat{\bar q}^a_{_I}$.
Thus there is only one independent
interference fragmentation function for $\pi\pi$ production, and it may be
factored out of the asymmetry,  $\sum_a
e_a^2\delta q_a\delta\hat{q}_{_I}^a=[{4/ 9}  (\delta u - \delta\bar u)- {1/
9} (\delta d - \delta\bar d)]\delta\hat{u}_{_I}$.
For the $K\overline{K}$ system, the isospin 
symmetry requires $\delta\hat{u}_{_I}=\delta\hat{d}_{_I}$, and $\delta\hat{s}_{_I}\ne 0$, 
implying  $\sum_a e_a^2\delta q_a\delta\hat{q}_{_I}^a=[{4/ 9}  (\delta u -
\delta\bar u)+ {1/ 9} (\delta d - \delta\bar d)]\delta\hat{u}_{_I}
+1/9 
(\delta s - \delta\bar s) \delta\hat{s}_{_I}$.  Similar application of
isospin symmetry and charge conjugation to the $\rho$ and $\sigma$ 
fragmentation functions that appear in the denominator of 
eq.\thinspace (\ref{asymmetry}) leads to a reduction in the number of 
independent functions:
$\delta\hat{u}_i
=\delta\hat{d}_i=\delta\hat{\bar u}_i=\delta\hat{\bar d}_i$ and 
$\delta\hat{s}_i=\delta\hat{\bar s}_i$
for $i=\{0,1\}$.
Finally we note that application of
the Schwartz inequality puts an upper bound on 
the interference fragmentation function,
$\delta\hat{q}_{_I}^2\leq 4\hat{q}_0\hat{q}_1/3$ for each flavor.

To summarize: Transversity is interesting, but hard to measure.  Two
soft hadronic processes must enter in the QCD analysis of a process
involving transverse spin in order to access $\delta q$.  After
reviewing previous suggestions, I have presented a new candidate:
observation of a $\vec k_1 \times \vec k_2 \cdot \vec S_\perp$
asymmetry in two meson production in the current fragmentation
region of DIS.  The analysis requires the study of twist-two
interference quark fragmentation functions in helicity density matrix
formalism.  The technique developed here can also be
applied to other processes. Straightforward applications include a
new asymmetry in two meson production from a longitudinally
polarized nucleon,\cite{ref2} and asymmetries in $e^+ e^-\rightarrow
(\pi^+\pi^- X) (\pi^+\pi^- X)$. A somewhat more complicated extension
can be made to two-meson production in single polarized
nucleon-nucleon collisions --- $p\vec p_\perp\rightarrow \pi^+\pi^-
X$, {\it etc.}  The last of these is of particular interest for the
polarized $pp$ collider at RHIC.  These applications will be presented
in future publications.

\vspace*{1cm}

I would particularly like to thank my collaborators, Xuemin Jin and
Jian Tang.  I am also grateful to John Collins and Xiangdong Ji
for conversations relating to this subject several years ago.  I 
would also like to thank the RIKEN-BNL Research Center where this
work was completed. This work is supported in part by funds
provided by the U.S. Department of Energy (D.O.E.) under cooperative 
research agreement \#DF-FC02-94ER40818, and by the RIKEN BNL Research Center.

\end{document}